# Slow scintillation time constants in NaI(Tl) for different interacting particles

C. Cuesta,[1] M.A. Oliván,[1] J. Amaré,[1] S. Cebrián,[1] E. García,[1] C. Ginestra,[1] M. Martínez,[1, a)] Y. Ortigoza,[1] A. Ortiz de Solórzano,[1] C. Pobes,[1] J. Puimedón,[1] M.L. Sarsa,[1, b)] J.A. Villar,[1] and P. Villar[1]

*Laboratorio de Física Nuclear y Astropartículas, Universidad de Zaragoza, Pedro Cerbuna 12, 50009 Zaragoza, SPAIN*
*Laboratorio Subterráneo de Canfranc, Paseo de los Ayerbe s.n., 22880 Canfranc Estación, Huesca, SPAIN*

Very large thallium doped sodium iodide crystals operated underground and in very low background environment in the context of a dark matter search experiment have been used to determine scintillation components in the tens of ms range in the light pulse induced by different interacting particles: $\gamma/\mu$ and $\alpha$.

## I. INTRODUCTION

Thallium doped sodium iodide scintillators have been widely used for radiation detection[1,2] since they were proposed by R. Hofstadter in 1948[3]. Because of their very high light yield, NaI(Tl) detectors became very soon one of the most convenient options for many applications, from radiology to environmental monitoring, as examples. NaI(Tl) detectors have other advantages, very large crystals can be grown and raw material is not expensive; also high efficiency for gamma ray detection is easily achieved. Nevertheless, NaI hygroscopic character makes difficult the manipulation of the crystals and its application in the very low energy x-ray regime. Main scintillation time constants of NaI(Tl) are well reported in the bibliography (see[1] and references therein): the dominant decay time of the scintillation pulses is in the range 230-250 ns[4–6], but slower components of $1.5\,\mu s$[4] and $0.15\,s$[7] have been also reported. Other possible phosphorescence components could be present at much longer timescales[8], but data are scarce. Differences in the scintillation time constants for different particles are also well known and have been used for discrimination purposes: alpha vs. gamma interactions at high energy[9–12] and nuclear recoils vs electron recoils at very low energies[13–17]. In the latter case, differences are quantified with an effective mean decay time, and are so small that only a statistical discrimination of events can be pursued. Dependence of the time constants with the energy of the particle has been also clearly established[14,17].

NaI(Tl) detectors have been used since the 90's in the search for the hypothetical Dark Matter (DM) particles filling our galactic halo and explaining an important part of the missing Universe mass[16,18–23]. Among these experiments, DAMA/LIBRA results have produced a large impact in the field by observing an annual modulation in the rates compatible with that expected for DM[23]. ANAIS (Annual modulation with NaI Scintillators) is an experiment to be carried out at the Laboratorio Sub-terráneo de Canfranc (LSC) in Spain, trying to confirm the presence of such a modulated signal using the same target and technique[24,25].

As a byproduct of the ANAIS prototype operation phase, scintillation decay time constants in several large NaI(Tl) detectors have been studied in different temporal scales and for different interacting particles, profiting from the very low event rate and stability of the experiment due to the underground operation at the LSC and the ultralow radioactive background environment. Up to our knowledge, it is the first time such a thorough study of the scintillation time constants in NaI(Tl) in the millisecond range has been reported.

## II. EXPERIMENTAL SET-UP

Four large NaI(Tl) crystals (see Tab. I) have been used to derive the results presented in this letter:

- a 10.7 kg hexagonal prism (distance between opposite vertices in the hexagonal face 15.94 cm, and 20.32 cm length) made by BICRON and stored underground at LSC since the late eighties. The original detector was opened at the University of Zaragoza (UZ) and re-encapsulated using OFHC copper;

- a 9.6 kg parallelepiped prism (10.16x10.16x25.40 cm³) grown by Saint Gobain Ltd. and encapsulated in ETP copper at the UZ after staying at a surface laboratory for several years, kept in dry atmosphere. Data used in this work were taken after almost three years of operation underground;

- two 12.5 kg cylindrical crystals (12.07 cm $\phi$, 29.85 cm length) grown by Alpha Spectra Inc. from the same ingot, using low potassium content selected NaI powder, and encapsulated in OFHC copper. Data taking started immediately after taking them underground.

Teflon tape (about 2 mm thick) plus a reflecting multilayer foil (3M Vikuiti$^{TM}$), both wrapping the NaI crystals, are used for light diffusion and reflection in the two former detectors and only Teflon in the latter two. Quartz optical windows (3" diameter) in both sides of the encapsulations allow the coupling of two photomultiplier tubes (PMTs) per crystal. In the data used along

---

a)Fundación ARAID, María de Luna 11, Edificio CEEI Aragón, 50018 Zaragoza, Spain
b)Electronic mail: mlsarsa@unizar.es; corresponding author

TABLE I. Mass, Manufacturer, PMT model used and name of the NaI(Tl) detectors used throughout this work.

| Name | mass | Manufacturer | PMT model |
|------|------|--------------|-----------|
| ANAIS-0 | 9.6 kg | Saint Gobain / UZ | R6956MOD |
| PIII | 10.7 kg | BICRON / UZ | R11065SEL |
| D0 | 12.5 kg | Alpha Spectra / UZ | R6956MOD |
| D1 | 12.5 kg | Alpha Spectra / UZ | R11065SEL |

TABLE II. Parameters of the digitizers used in this work.

| Digitizer | sampling rate | window | resolution |
|-----------|---------------|--------|------------|
| MATACQ | 2 GS/s | 1.25 $\mu$s | 12 bits |
| TDS5034B-207 | 25 MS/s | 320 ms | 8 bits |
| TDS5034B-208 | 250 MS/s | 40 $\mu$s | 8 bits |

this work Hamamatsu low background bialkali PMTs have been used (models R11065SEL and R6956MOD), see Tab. I. Coupling of the PMTs to the quartz windows has been done for all the modules at the LSC. According to the manufacturer, for both models maximum spectral response should be found at 420 nm wavelength, matching properly the emission of NaI(Tl), being the sensitivity range 200 - 650 nm and 300 - 650 nm, respectively. Nominal quantum efficiencies at 420 nm (from provider) are 29 and 33% for the R11065SEL, and 34 and 35% for the R6956MOD PMT units used in this work.

Detectors were operated inside a shielding consisting of 10 cm archaeological lead and 20 cm low activity lead, all enclosed in a PVC box tightly closed and continuously flushed with boil-off nitrogen. Two detectors could be operated at the same time. Three plastic scintillators (0.5 m$^2$ each) were mounted on top of the shielding to tag muons. The LSC is below 850 m rock overburden (equivalent to 2450 m.w.e.), implying a reduction on the muon flux of about four orders of magnitude with respect to that found at the surface[26].

No radioactive sources have been required for this study. Alpha particle events used have internal origin: uranium and thorium natural chains isotopes are present in small quantities in the bulk of the NaI(Tl) crystals and originate homogeneously distributed events populations[25]. Gamma and muon events from radioactive environmental background have been also used throughout this work.

The main part of the electronic chain used for the data taking was designed for the search of the very low WIMP energy depositions and its complete description is not required in this letter, we will only refer below to the digitization stage using a MATACQ based VME board (2 GS/s sampling rate, 2520 sampled points, and 12 bits of vertical resolution). Data triggering is done in logical OR mode between different detectors operated at the same time and in logical AND between the two PMT signals from each detector. In the dark matter acquisition mode, triggering at photoelectron level in each PMT is accomplished, but for data shown hereafter, dynamic range was tuned to register events in the MeV range (in electron equivalent energy, MeVee) by conveniently attenuating the signals. Events triggering the plastic muon vetoes reset a counter that is read after every NaI(Tl) event and saved with it, allowing to tag muon related events by off-line analysis.

Data acquisition in the very long timeline basis required the design of an special branch in the electronic chain. It consists of two Tektronix phosphor oscilloscopes (TDS5034B) in two different time scales and also different vertical ranges, having 8 bits of vertical resolution. Main characteristics of the oscilloscopes used in this work are: 350 MHz bandwidth, up to 5 GS/s sampling rate capability, 4 channels and between 500 and 8·10$^6$ sampled points. Trigger was done independently for the Tektronix oscilloscopes and the MATACQ board. The former were only triggered by very high energetic events (above 2 MeVee) and measurements were done for one detector each time. The most relevant configuration parameters of the two used oscilloscopes and the MATACQ board are summarized in Tab. II. TDS5034B-207 dynamic range allows to identify individual photoelectrons in the pulse tail, whereas TDS5034B-208 measures the whole of the pulse, whose area is taken as energy estimator.

## III. RESULTS

### A. Events selection

Muon events can only be tagged by the plastic scintillators veto signals for the MATACQ data (shortest timescale). For the oscilloscopes data, muon and gamma events are considered the same population; however, we can expect to find more muons than gammas above the $^{208}$Tl line at 2614.6 keV. Muons are found at a rate of about 45 events/day whereas for alpha events the corresponding rate is between 800 and 3000 events/day, depending on the crystal bulk contamination in uranium and thorium chains. $\alpha$ are separated from $\gamma/\mu$ events by a very simple cut based on the different relationship between amplitude and area of the fast component of the pulse (1 $\mu$s window), hereafter called fast pulse. Two populations can be clearly distinguished, estimating as 100% the efficiency of the discrimination above 2 MeVee and up to the saturation of the PMT signal. No assumptions about the timing constants are required to apply such a discrimination. Neither a difference between $\gamma$ and $\mu$ events has been found, as expected, nor a significant difference between fast pulses coming from the four studied crystals. Because of that, in sec. III B and III C only data from ANAIS-0 will be shown.

### B. Pulses in the microsecond range

MATACQ data corresponding to ANAIS-0 detector have been selected in a region of interest (ROI) corresponding to about 2.5 to 3 MeVee (4.2 to 5 MeV for $\alpha$ particles, because the quenching factor has been measured to be 0.6 at these energies). We have built average pulses for the three populations of events corresponding to interactions of $\alpha$, $\mu$ and $\gamma$ in the crystal, and normalized to the same pulse area. The corresponding pulses can be seen in Fig. 1. No difference between gamma and muon events is noticed, and in the following we will consider both as a single population, as far as a different behavior in the longer timescales is not expected.

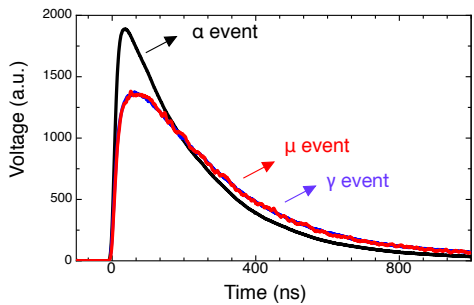

FIG. 1. Average pulses from $\alpha$ events and those produced by $\gamma$ and $\mu$, normalized to the same pulse area.

TABLE III. Results of the fits from 0 to 1000 ns after the pulse onset for alpha and gamma/muon events.

| Fit parameter | $\alpha$ events | $\gamma/\mu$ events |
|---|---|---|
| $\tau_{rise}$ (ns) | $17.98 \pm 0.12$ | $28.23 \pm 0.14$ |
| $\tau_{decay}$ (ns) | $219.28 \pm 0.52$ | $287.35 \pm 0.51$ |
| $A_{rise}/A_{decay}$ | $1.223 \pm 0.004$ | $1.123 \pm 0.003$ |

Pulses have been fitted to a combination of two exponential decays ($A_i e^{-t/\tau_i}$), independently for $\alpha$ and $\gamma/\mu$ events: one corresponds to the rise of the pulse, and the second, to the decay. However, the presence of slower components could affect the results of the fit in this region (see sec. III C). Results are given in Tab. III. In the case of $\alpha$ pulses, rise time is compatible with contributions coming from light propagation in the crystal and width of Single Electron Response (SER) of the PMT (gaussian shaped with FWHM of 12 ns for the R6956 PMT), hinting at a prompt light emission. Decay time for $\alpha$ pulses is compatible with that reported in the bibliography. However, for the $\gamma/\mu$ pulse, some more complicated scintillation mechanism has to be coming into play: delayed scintillation by an intermediate non-scintillating excited state could be responsible of the slower rise time; moreover, we obtain a decay time slightly slower than most of the values reported in the bibliography, although it is worth remarking that in those works only a mean time of the pulse, instead of a decay constant, is given. Ref. [15] also supports slower time constants, although the study is done at much lower energies.

### C. Pulses in the 40 microseconds range

Scope data corresponding to ANAIS-0 detector have been used to build average pulses, normalized to the same fast pulse area, for $\alpha$ and $\gamma/\mu$ populations in the same energy region considered in sec. III B. Important effects of the RC time constant of the PMT readout circuit can be observed in Fig. 2, limiting the conclusions derived for the possible presence of additional scintillation time constants in the few microseconds range. Clearly pulses undershoot the baseline, being not recovered in the 40 $\mu$s studied range, but in the first milliseconds. However, it can be noticed a much more important undershooting for $\alpha$ events than for those attributable to $\gamma/\mu$, pointing at a possible additional slow scintillation component in the latter that could partially compensate the undershooting. Similar effects have been observed for all the detectors studied and a modification of the PMT readout circuit would be required to further clarify this issue.

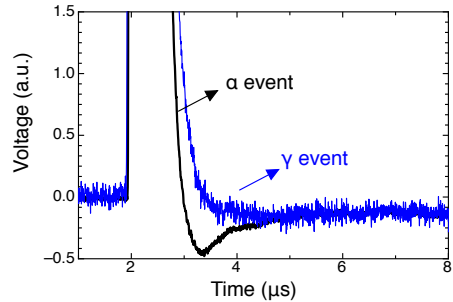

FIG. 2. Pulses from $\alpha$ and $\gamma/\mu$ events in a zoomed view of the pulse baseline to remark the pulse undershoot.

### D. Pulses in the 320 milliseconds range

For the longest timescale, a different approach has been followed: photoelectrons (ph.e.) have been identified individually, at a given position of the pulse, and an histogram has been produced with the descending temporal distribution for all the events in the ROI, separately for $\alpha$ and $\gamma/\mu$ events for each of the four NaI(Tl) detectors studied. The corresponding distributions can be seen in Fig. 3, conveniently normalized (although independently for each detector) to the same fast pulse area and averaged according to the number of events in the ROI for each population. Similar ROI have been studied for ANAIS-0 (see sec. III B), D0 and D1 detectors, but PIII ROI corresponds to higher energies (from 5 to 5.3 MeVee). Selection of individual ph.e. is done very efficiently in the MATACQ data by applying a peak search algorithm, however, in the data from TDS5034B-207, we expect many ph.e. to be lost because of the sampling rate used, and also we expect differences between detectors because of the different SER of each PMT model used; therefore, only relative values concerning the ph.e. number in the slow scintillation components for each detector are considered and comparison with the fast component, or among detectors, is meaningless.

In Tab. IV the ratio between the average ph.e. number for $\alpha$ and $\gamma/\mu$ events is shown for the four detectors studied, as well as the mean time, calculated only with the tail: from 4 till 304 ms after the pulse onset. The most relevant difference is the total number of ph.e. identified in these slow components: $\gamma$ and $\mu$ events excite much more efficiently the long-lived states contributing to them, whereas similar mean time constants are found for both types of interacting particles, being in all cases $\alpha$ events slightly faster. Because of the quenching factor for $\alpha$ vs $\beta/\gamma$ events in NaI(Tl) and the fact that we have normalized to the fast pulse area and not to the energy deposited, the referred effect is still much more important in terms of photoelectron production per unit of energy. In spite of these similarities, significant differences among the studied crystals can be reported; hence, impurities or defects could play a role in this slow scintillation mechanism.

TABLE IV. Ratio between the average ph.e. number for $\alpha$ and $\gamma/\mu$ events and mean pulse time calculated from 4 till 304 ms after the pulse onset.

| Detector | $N_\alpha/N_{\gamma/\mu}$ | $\tau_{\alpha,mean}$ (ms) | $\tau_{\gamma/\mu,mean}$ (ms) |
|---|---|---|---|
| ANAIS-0 | $0.42 \pm 0.15$ | 75.48 | 80.25 |
| PIII | $0.81 \pm 0.21$ | 68.33 | 68.99 |
| D0 | $0.48 \pm 0.11$ | 92.12 | 101.72 |
| D1 | $0.44 \pm 0.06$ | 88.25 | 99.51 |

TABLE V. Results of the fit to two exponential decays from 4 to 304 ms after the pulse onset for $\alpha$ and $\gamma/\mu$ events.

| Detector | Particle | $\tau_1$ (ms) | $\tau_2$ (ms) | $A_1/A_2$ |
|---|---|---|---|---|
| ANAIS-0 | $\alpha$ | $18.5 \pm 0.6$ | $83.3 \pm 0.5$ | $0.43 \pm 0.01$ |
| | $\gamma/\mu$ | $29.3 \pm 3.5$ | $94.9 \pm 2.4$ | $0.39 \pm 0.02$ |
| PIII | $\alpha$ | - | $62.0 \pm 0.9$ | - |
| | $\gamma/\mu$ | - | $69.2 \pm 1.5$ | - |
| D0 | $\alpha$ | $54.6 \pm 3.7$ | $180.7 \pm 18.6$ | $1.82 \pm 0.38$ |
| | $\gamma/\mu$ | $52.5 \pm 2.4$ | $171.5 \pm 4.2$ | $0.66 \pm 0.04$ |
| D1 | $\alpha$ | $42.2 \pm 7.2$ | $128.6 \pm 12.8$ | $0.87 \pm 0.29$ |
| | $\gamma/\mu$ | $61.5 \pm 3.2$ | $191.4 \pm 10.9$ | $1.21 \pm 0.08$ |

Fits have been carried out for $\alpha$ and $\gamma/\mu$ events for the four detectors studied, using the range from 4 to 304 ms after the onset of the pulse, trying to distinguish different scintillation components. Two exponential decays have been considered in the fit, except in the case of PIII data (affected by a poor statistics and a higher energy of the selected events). Corresponding results are summarized in Tab. V. No clear conclusion can be drawn from the fits.

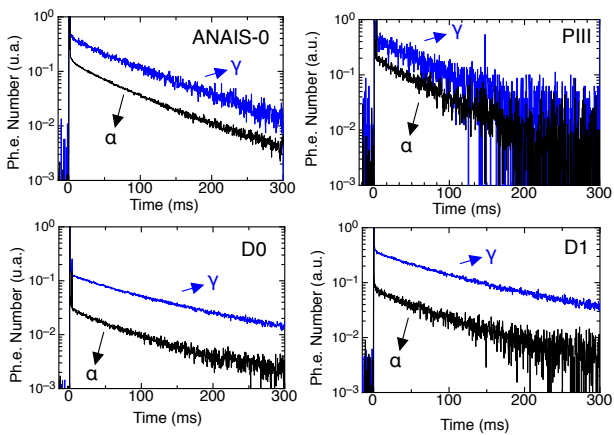

FIG. 3. Ph.e. distribution from $\alpha$ and $\gamma/\mu$ events in a 320 ms temporal scale, normalized to the same fast pulse area.

## IV. CONCLUSIONS

Scintillation time constants in NaI(Tl) crystals from ns up to 300 ms have been studied for $\alpha$ as well as for $\gamma/\mu$ events in the high energy regime. Very different behavior for both kind of interactions is confirmed, specially in the capability of exciting the slow scintillation mechanism (70-100 ms mean time constants). Significant differences have been observed among results derived from different NaI(Tl) crystals, pointing at an origin of such scintillation related to impurities or defects more than to the thallium doping. However, all the crystals studied showed such effects and further work is in progress to determine how this slow scintillation could affect the application of these detectors in DM searches.

## ACKNOWLEDGMENTS

This work has been financially supported by the Spanish and European Regional Development Fund MINECO-FEDER under grant FPA2011-23749, Consolider-Ingenio 2010 Programme under grants MultiDark CSD2009-00064 and CPAN CSD2007-00042 and the Gobierno de Aragón.

[1] J.B. Birks, *The theory and practice of scintillation counting*. Pergamon Press Ltd., 1964.
[2] P. Lecoq, A. Annenkov, A. Gektin, M. Korzhik, C. Pedrini, *Inorganic scintillators for detector systems* (Springer-Verlag, 2006)
[3] R. Hofstadter, *Phys. Rev.* **75** (1949) 796.
[4] J.C. Robertson and J.G. Lynch, *Proc. Phys. Soc.* **77** (1961) 751.
[5] F.S. Eby and W.K. Jentschke, *Phys. Rev* **96** (1954) 911.
[6] J.S. Schweitzer and W. Ziehl, *IEEE Trans. Nucl. Sci.* **30** (1983) 380.
[7] S. Koićki, A. Koićki, and V. Ajdačič, *Nucl. Instrum. Meth.* **108** (1973) 297.
[8] C.R. Emigh and L.R. Megill, *Phys. Rev.* **93** (1954) 1190.
[9] P. Doll, *et al.*, *Nucl. Instrum. Meth. A* **285** (1989) 464.
[10] J.C. Barton, *Appl. Rad. Isot.* **47** (1996) 997.
[11] K. Ichihara, *et al.*, *Nucl. Instrum. Meth. A* **515** (2003) 651.
[12] C. Bacci, *et al.*, *Phys. Lett. B* **293** (1992) 460.
   R. Bernabei, *et al.*, *Nucl. Instrum. Meth. A* **592** (2008) 297.
[13] R. Bernabei, *et al.*, *Phys. Lett. B* **389** (1996) 757.
[14] D.R. Tovey, *et al.*, *Phys. Lett. B* **433** (1998) 150.
[15] V. Kudryavtsev, *et al.*, *Phys. Lett. B* **452** (1999) 167.
[16] G. Gerbier, *et al.*, *Astropart. Phys.* **11** (1999) 287.
[17] L. Miramonti, *Rad. Phys. Chem.* **64** (2002) 337.
[18] K. Fushimi, *et al.*, *Phys. Rev. C* **47** (1993) 425.
   K. Fushimi, *et al.*, *Astropart. Phys.* **12** (1999) 185.
[19] M.L. Sarsa, *et al.*, *Phys. Lett. B* **386** (1996) 458.
   M.L. Sarsa, *et al.*, *Phys. Rev. D* **56** (1997) 1856.
[20] R. Bernabei, *et al.*, *Phys. Lett. B* **450** (1999) 448.
   R. Bernabei, *et al.*, *Riv. Nuovo Cim.* **26** (2003) 1.
[21] G.J. Alner, *et al.*, *Phys. Lett. B* **616** (2005) 17.
[22] K. Fushimi, *et al.*, *J. Phys.: Conf. Ser.* **203** (2010) 012046.
[23] R. Bernabei, *et al.*, *Eur. Phys. J. C* **56** (2008) 333.
   R. Bernabei, *et al.*, *Eur. Phys. J. C* **67** (2010) 39.
[24] C. Pobes, *et al.*, *J. Phys.: Conf. Ser.* **203** (2010) 012044.1-3.
   J. Amaré *et al.*, *J. Phys.: Conf. Ser.* **375** (2012) 012026.
[25] S. Cebrián, *et al.*, *Astropart. Phys.* **37** (2012) 60.
[26] E.V. Bugaev, *et al.*, *Phys. Rev. D* **58** (1998) 054001.